# Time-reversed magnetically controlled perturbation (TRMCP) optical focusing inside scattering media

Zhipeng Yu[2,4,#], Jiangtao Huangfu[3,#], Fangyuan Zhao[1, 5], Meiyun Xia[1, 5], Xi Wu[3], Xufeng Niu[1, 5], Deyu Li[1, 5], Puxiang Lai[2,4,*], and Daifa Wang [1,5,*]

[1]School of Biological Science and Medical Engineering, Beihang University, Beijing, China, 100083

[2]Department of Biomedical Engineering, Hong Kong Polytechnic University, Hong Kong

[3]Laboratory of Applied Research on Electromagnetics (ARE), Zhejiang University, Hangzhou, China, 310027

[4]Shenzhen Research Institute, Hong Kong Polytechnic University, Shenzhen, China, 518057

[5]Beijing Advanced Innovation Centre for Biomedical Engineering, Beihang University, Beijing, China，102402

*Co-corresponding authors: daifa.wang@buaa.edu.cn; puxiang.lai@polyu.edu.hk

[#]These authors contribute equally to this work.

**Abstract**: Manipulating and focusing light deep inside biological tissue and tissue-like complex media has been desired for long yet considered challenging. One feasible strategy is through optical wavefront engineering, where the optical scattering-induced phase distortions are time reversed or pre-compensated so that photons travel along different optical paths interfere constructively at the targeted position within a scattering medium. To define the targeted position, an internal guidestar is needed to guide or provide a feedback for wavefront engineering. It could be injected or embedded probes such as fluorescence or nonlinear microspheres, ultrasonic modulation, as well as absorption perturbation. Here we propose to use a magnetically controlled optical absorbing microsphere as the internal guidestar. Using a digital optical phase conjugation system, we obtained sharp optical focusing within scattering media through time-reversing the scattered light perturbed by the magnetic microsphere. Since the object is magnetically controlled, dynamic optical focusing is allowed with a relatively large field-of-view by scanning the magnetic field externally. Moreover, the magnetic microsphere can be packaged with an organic membrane, using biological or chemical means to serve as a carrier. Therefore, the technique may find particular applications for enhanced targeted drug delivery, and imaging and photoablation of angiogenic vessels in tumours.

**Introduction**

Due to the strong scattering of light in biological tissue, conventional optical manipulation methods, such as using objective lenses, can only focus visible and near infrared (NIR) light to shallow depths of up to a few hundred micrometers[1]. With the increase of propagation distance, the number of ballistic photons decays exponentially and becomes obsolete beyond one transport mean free path ($l_s$, ~1 mm for human skin).

However, unless being absorbed, these photons do not disappear; they just become diffusive photons that travel along random optical paths, forming speckle patterns provided a sufficiently long coherence length of light[1]. Noninvasive optical focusing beyond $l_s$ (also referred to as the optical diffusion limit) has been desired yet considered challenging for long[2]. Until recently, researchers started to notice that the seemingly random scattering events and the resultant speckles are actually deterministic within the speckle decorrelation window[3,4]. Since then, various wavefront engineering approaches have been developed[5-9] to compensate for or reverse the scattering-induced phase distortions, so that diffused light travelling along different optical paths may interfere constructively again at the targeted position, forming an optical focus out of the seemingly random speckle background.

Current wavefront engineering approaches mainly include two categories, i.e. optical wavefront shaping techniques[10-15] and optical phase conjugation[7,16,17]. While related, these two categories are different in the principle of implementation. In iterative wavefront shaping, a spatial light modulator (SLM) is used to shape the spatial phase distribution of an incident beam before it enters a scattering medium. The optimum phase pattern is obtained by an iterative algorithm [10,11] that optimizes a feedback signal, or by measuring the transmission matrix of the scattering medium[14,18,19]. Whilst effective in generating an intense optical focus with a high peak-to-background ratio (PBR) and different scales of resolution (depending on the nature the guidestar), wavefront shaping-based implementations are usually time-consuming due to the requirement of thousands of phase pattern updates and feedback signal measurements[10,19]. Focusing by optical phase conjugation (OPC), in comparison, does not involve that many iterations, and hence can be much faster. In OPC, distorted light exiting the scattering medium is holographically recorded by using a phase conjugation mirror (PCM), usually a photorefractive material[7,20] or a well-aligned digital camera-SLM module[16,17,21]. Reading out the hologram generates a phase conjugated copy of the original signal beam, which will travel, albeit tortuously, back to the scattering medium and converge to the point of origin. Within a complete OPC operation cycle, up to four interfering patterns are recorded, and the phase pattern on the SLM needs to be refreshed for once. Currently, OPC-based schemes have enabled optical focusing within several ms[22,23], being inherently much faster than their iterative peers.

To enable focusing inside a scattering medium, which is more of biomedical interests, an internal guidestar is required to modulate diffused light traversing the voxel of interest or to produce an effective light source (real or virtual) within the medium. It could be embedded probes, such as fluorescent[16,17,24] or nonlinear beads[25]. Their embedding procedures, however, are usually invasive and the resultant focal positions are fixed and limited. Ultrasound, on the other hand, scatters ~1000 times weaker than light does and penetrates deep (several or tens of centimeters) into tissue. Moreover, diffused light can be spectrally encoded by ultrasound locally and can be converted into weakly scattered ultrasound. Therefore, ultrasonic mediation has served as an encouraging noninvasive internal guidestar in a class of techniques recently developed by researchers, such as time-reversed ultrasonically encoded (TRUE) optical focusing[6,26-28], time reversal of variance-encoded (TROVE) light[29], and photoacoustically guided wavefront shaping (PAWS)[10,14]. More recently, inspired by the employment of ultrasound, moving absorbers[30,31] and microbubble collapsing induced optical perturbations[32] have also been introduced as the internal guidestar.

While promising, the existing internal guidestars are limited and not yet suitable for broad applications. For example, the efficiency of ultrasonic modulation is usually low[6,26-28], the ultrasonic guidestar-based approaches require physical contact (e.g. with water) for acoustic coupling and reduced loss at high acoustic frequencies, and the adaptive perturbation-based methods lack specific recognition or control[30,31]. To tackle the aforementioned limitations, in this paper we propose a new approach called time-reversed magnetically controlled perturbation (TRMCP) optical focusing, using magnetically controlled optical absorbing microspheres as internal guidestar for digital optical phase conjugation (DOPC). We show that optical focusing within scattering media can be achieved by time-reversing the scattered light perturbed by the magnetic microsphere. Since the object is accurately positioned and controlled by a magnetron device, dynamic optical focusing is allowed with a relatively large field-of-view by scanning the magnetic field externally. Such focusing may potentially benefit a wide range of biomedical applications in vessel-like aquatic environment, such as blood vessels and lymph vessels.

## Methods

1. Experimental setup

The experimental setup of TRMCP is showed in Fig. 1(a). A 532nm continuous laser

(EXLSR-532-200-CDRH, SPECTRA PHYSICS) is used as the light source, and its coherence length is 300 m. The laser output is split into two arms, a sample beam and a reference/playback beam, by a beam splitter ($BS_1$). In the hologram recording stage, the sample beam is expanded and illuminates the front surface of a scattering medium composed of two diffusive Scotch tapes (3M, ~60 μm thick) separated by 5.5 cm. Distorted sample light exiting the medium is collected and relayed to a scientific CMOS camera (sCMOS, pco.edge 5.5, PCO), where it interferes with the reference beam. The interfering patterns are transferred to the computer to compute the optical field, whose conjugation is then transferred to the spatial light modulator (SLM, PLUTO-VIS-056, HOLOEYE). In the hologram readout stage, the sample beam is blocked, and a playback beam illuminates the SLM, generating a phase conjugated copy of the original sample beam, which travels back to the scattering medium and converges to the position of incidence at the front surface of the first scattering layer ($S_1$), assuming there is no internal guidestar between the scattering layers. In additional to abovementioned components, a single-mode fiber (SMF) and lens $L_1$ are used to shape the reference beam (planar, 25.4 mm in diameter), lens $L_2$ is used to collect light exiting the scattering medium and adjust the speckle grain size in $sCMOS_1$, and lens $L_3$ is positioned in front of $sCMOS_1$ to image the surface of SLM or Mirror M. Four fast shutters ($FS_{1-4}$) are used to control the on and off of different optical beams. Between the two layers of diffusive tapes, a capillary tube is used to mimic a vessel. By changing the position of the magnetic microspheres within the capillary tube under the magnetic field (described in next section), distribution of scattered light is tuned and recorded. Another camera (CMOS camera) positioned conjugated to the magnetic microsphere plane is used to observe the time-reversed sample beam. Lastly, it should be noted that with the 5.5 cm distance between the two Scotch tapes in our setup, few ballistic photons are collected and sent to the sCMOS camera for hologram recording[31].

## 2. Magnetic control system

A magnetic system is built to accurately control the positioning and movement of the magnetic microsphere within the tube, as shown in Fig. 2. A custom-designed electromagnet is mounted on a 2-D translation stage, and a sharp needle is adhered onto the upper surface of the magnet to generate a narrow magnetic field. The diameter of the needle tip is 0.05 mm, with a taper angle of 3.0 degree and a length of 29 mm. In our experiment, a square capillary (inner cross-section 0.3×0.3 mm$^2$, wall thickness 0.1 mm) is used to mimic a vessel and placed 0.1 mm above the tip of needle. With this setup, a magnetic microsphere inside the tube can be attracted to the peak magnetic intensity position. Consequently, by scanning the translation stage along the X direction, accurate positioning, moving, as well as monitoring of the magnetic microsphere inside the tube is achieved. To avoid light being blocked by the magnetic needle, the magnetic needle is moved out of the field of view along the Z direction, when the light field information is under acquisition. It should be noted that during this moving out process, the magnetic field is varied, which may cause some microsphere perturbation (up to ~100 μm for the used ~200 μm diameter sized microsphere). To avoid this effect, during the needle moving in/out procedures, the electrical current through the electromagnet coil is shut off. As the needle and the electromagnet core are made of soft magnetic material, silicon steel, the remnant magnetism is negligible in current-free condition.

Herein, a gauss meter (CH-3600, CH-HALL, China) is used to measure the magnetic intensity distribution in a vertical plane above the needle. The probe diameter of the gauss meter is 1 mm. As we can see from Fig. 2, the magnetic intensity profile in the microsphere position seems not so sharp. It is largely caused by the relatively large size of the probe. The accuracy of repeated positioning of the magnetic control system is also quantified. A microsphere with a diameter of 200 μm is controlled to move between two positions (with an inter-distance of 200 μm) for 10 times along the X direction. As a result, the positions produce a standard deviation of 9 μm and a maximum deviation of 19 μm. Thus, the open-loop positioning of our magnetic control system is sufficiently accurate, with a position standard deviation of within 5% of the object size.

### 3. Principle of TRMCP optical focusing

The focusing procedure is divided into two stages, i.e., phase recording and playback, respectively. In the phase recording stage, the magnetic microsphere (target absorption $P(r, t_i)$, $i=1, 2$) is first positioned at Location 1 (Fig 3a). The resultant light fields are $U_1$ and $E_1$ at the microsphere plane and the camera plane, respectively. They are associated through the transmission matrix ($T_1$) of the system between the microsphere plane and the camera plane, thus $E_1=T_1 \cdot U_1$. When the magnetic microsphere is moved to Location 2 (Fig. 3b), the light fields become $U_2$ and $E_2$, respectively, which are associated with $E_2=T_2 \cdot U_2$. Assuming the system is linear and is sufficiently stable within the focusing procedure, $T_1=T_2=T$ and the differences of two sets of light field $\Delta U=U_1-U_2$ and $\Delta E=E_1-E_2$ caused by the magnetic microsphere perturbation can be expressed by $\Delta E=T\cdot\Delta U$. In the playback stage (Fig. 3c), a phase pattern $\Delta E^*$ that is conjugated to $\Delta E$ is displayed on the SLM. The reference beam now serves as a playback beam that is modulated and reflected by the SLM surface. Due to the nature of phase conjugation, the reflection travels back to the scattering sample, although tortuously, and reaches the magnetic plane, resulting in an optical field of $U_3$ expressed by

$$U_3=T^+E^*=T^+(T\Delta U)^*=(T^\dagger T\Delta U)^*\approx(\Delta U)^* \qquad (1)$$

where $^*$ denotes a complex conjugate. $+$ and $\dagger$ denotes a transpose and a conjugate transpose. Approximately, $T^\dagger T\approx I$ ($I$ is the identity matrix) assuming the system is time invariant during the process. As seen from the equation, $U_3$ is the conjugate to the light field difference at the microsphere plane, and hence converges to Positions 1 and 2. That is, the time-reversed light converges onto the magnetically guided moving microsphere.

## 4. Characterization of the DOPC system

To successfully perform TRMCP focusing, the quality of the DOPC system must be ensured. For example, the pixels between the sCMOS camera and the SLM must be precisely aligned and matched. In our experiment, a specific pattern is displayed on the SLM, then we carefully align the SLM and the sCMOS camera to ensure the pixel mismatch to be within one pixel. Moreover, a four-phase method is used to compensate for the SLM curvature-induced modulation error. As shown in Fig. 1, the SLM and the mirror M after FS$_3$ is adjusted to be perfectly perpendicular to the reference beam by using a retroflector, and the sCMOS camera is used to obtain the interference pattern of lights reflected by the SLM and mirror. We carefully adjust the inter-distance between L$_2$ and S$_2$ (S$_1$ removed in advance), as well as between L$_2$ and the sCMOS camera, so that a fully developed speckle patterns with appropriate speckle grain size are obtained in experiment. With the aforementioned optimization, an optical focus with a PBR up to 9,000 is obtained when the computed phase modulation pattern is displayed on the SLM (Fig. 4b). The full width at half maximum (FWHM) of the focus is 10μm and 7.5μm, respectively, along the lateral directions. In contrast, a random speckle pattern is obtained when a uniform or random phase pattern is displayed on the SLM (Fig. 4a). theoretically the PBR is $\sim\frac{\pi}{4}N$ for a phase-only SLM[33]. Here, $N$ is the number of speckle grains (controlled modes) on the SLM plane. In the experiment, each speckle grain occupies about 6×6 pixels on the SLM. So, the theoretical PBR is ~45,000. Therefore, the focusing efficiency (η) defined by the ratio between the measured and theoretical PBR values is ~0.2. This is well below unity yet considered experimentally reasonable, if the influence due to the imperfection in the alignment, curvature compensation as well as system stability is taken into account.

**Results and Discussion**

With a characterized and optimized DOPC system, experiments are performed to demonstrate the feasibility and performance of TRMCP optical focusing. As shown in Fig. 5a, magnetic particles sized from 20-50 μm are enveloped with opaque levorotatory polylactic acid (PLLA) to form a magnetic-organic compound[34].

The procedure is as follows: First, the magnetic microspheres and PLLA solution are mixed sufficiently and then emulsified. Second, the mixtures are poured (at a suitable speed) into a polyving akohol solution that has been put on a rotating centrifuges stage in advance. As a result, the magnetic microspheres are enveloped with PLLA films, and compound microspheres with different sizes are formed. At last, microspheres of specific size are obtained through filtering the solution using appropriate meshes. A camera is used to record the resultant speckle pattern at the microsphere plane in advance, as shown in Fig. 5b and Fig. 5c. One speckle grain occupies ~25 pixels on the camera, and the camera pixel size is 2.5×2.5 μm². The diameter of the microsphere range is ~200 μm (5000 pixels), encompassing ~200 speckle grains.

To mimic practical vessel application scenario, a square capillary with an inner cross-section of 0.30×0.30 mm² is used as the flowing channel for the magnetic microsphere. The distance between $S_1$ and $S_2$ is 5.5 cm, and the tube is positioned 3 cm away from $S_2$ (Fig. 1). In the experiment, the magnetic microsphere at first is away from the field of view, then is controlled to move step by step across the probed region. At each position, one hologram is recorded and the TRMCP is executed sequentially. Fig. 6a shows an example of focusing light to three adjacent positions by using TRMCP, where the magnetic microsphere is first placed at Location 1, then is controlled to move to Location 2, then to Location 3. Through light reflected by $BS_4$ in Fig. 1, the position variation of the microsphere is also monitored in real time by a digital microscope, as shown in Figs. 6 b-d. As seen, the microscopic microsphere images are consistent with the three time-reversed optical focal spots in terms of position and dimension. It is evident that the phase conjugated beam has successfully converged back to the perturbation origins, which confirms the feasibility of TRMCP focusing. However, it should be noted that intensities of the multiple speckle grains within each focus are not evenly distributed, which is probably due to the imperfection in the sample beam illumination, the compensated phase pattern retrieval, and the system stability.

A whole focusing cycle, from the onset of recording the first hologram to the playback of the time-reversed light, took about 4 seconds, which is sufficiently fast for *ex vivo* and phantom-based studies. For *in vivo* applications, the speed of the focusing process must be considerably reduced since seconds is usually longer than the optical decorrelation time (on the order of milliseconds) associated with physiological motions such as blood flow and aspiration in living biological tissue. The slow process in the current system is largely due to the use of a personal computer as the controller for data transfer and processing, the lagged response of the SLM, the mechanical scanning of the magnetic field (~1 s), the move in/out of magnetic needle (~1 s total), and the response time of electromagnet (0.4 s). To accelerate the focusing and make it suitable for living tissue applications where optical field decorrelates on the order of milliseconds due to physiological activities such as blood flow and respiration[35,36], further development will apply strategies to both the DOPC module and the magnetic controlling module. To build a faster DOPC system, a field programmable gate array (FPGA) combining with digital mirror device (DMD), a micro-electro-mechanical system (MEMS)-based SLM or a ferroelectric liquid crystal-based SLM will be considered[22,23,37].

Three more aspects also need to be addressed. First, the PBR of the optical focusing with magnetic microsphere in experiment was 32, while theoretically it can reach 112.5 according to $PBR_{TRMCP} = \frac{\pi}{4}\frac{N}{K}$ for a phase-only SLM[30], where $K$ is the number of the speckle grain on the microsphere plane of the sample beam. The above estimation contains a $\pi/4$ factor because the SLM used in experiment is a phase-only modulator. The efficiency, defined by the ratio between the measured and theoretical PBRs is only 28.4%, possibly due to factors such as (1) the boundary or mismatch between the tube and the phantom medium, as well as the rectangular shape of the tube (that introduce extra deflection to the sample beam; (2) the thickness of the tube, and (3) the alignment imperfection of the DOPC system. Second, in experiment the magnetic microsphere was only 0.1 mm above the tip, which is a close distance to ensure the movement resolution of the particle. To translate the system towards real applications, a customized high-speed mechanical scanning system and quickly responding electromagnet will be applied. Lastly but not the least, the medium flow within the tube may induce position variation to the magnetic microsphere, especially when the magnetic binding force is not sufficiently large. Such influence cannot be ignored in applications in vessel-like environment. A fast system discussed above can help to overcome the challenge, and furtherly, one may be able to temporarily block or slow down the flow (e.g. blood flow) by using clamps or rubber bands[6,38]. With the aforementioned improvement, TRMCP can potentially be useful for many applications in vessel-like aquatic environment, such as lymph vessels and blood vessels.

**Conclusion**

Optical focusing beyond the diffusion limit inside biological tissue has been long desired yet considered challenging. Aiming towards this goal, we propose to use a magnetically controlled microsphere movement that generates optical perturbation as internal guidestar for optical phase conjugation. The feasibility of this approach, referred to as time-reversed magnetically controlled perturbation (TRMCP) optical focusing, has been demonstrated experimentally herein and recently by Ruan *et al*. in an another independent study[39]. Compared to previous schemes in the field, this new approach has advantages in the playback efficiency and the guidestar scanning controllability, making it a promising solution to focus light into vessel-like medium at depths in tissue. Moreover, the magnetic microsphere can be packaged with an organic membrane, using biological or chemical means, and serve as a carrier. Therefore, this approach, once further engineered, may potentially find some applications for precisely targeted drug delivery, tumor angiogenic vessel imaging and photoablation.

**Acknowledgement**

The work has been supported by the National Natural Science Foundation of China (Nos. 61675013, 81671726, 617611660 and 81627805) and the Hong Kong Research Grant Council (No. 252044/16E).

## Author contributions

D. Wang, J. Huangfu, and P. Lai conceived the idea. Z. Yu, D. Wang, J. Huangfu, and P. Lai designed the system. Z. Yu, F. Zhao, M. Xia, X. Wu, D. Li, and X. Niu ran the experiments. Z. Yu, D. Wang, and P. Lai prepared the manuscript. D. Wang and P. Lai co-supervised the project. All authors were involved in the analysis of the results and manuscript revision.

## Additional information

Competing financial interests: The authors declare no competing financial interests.

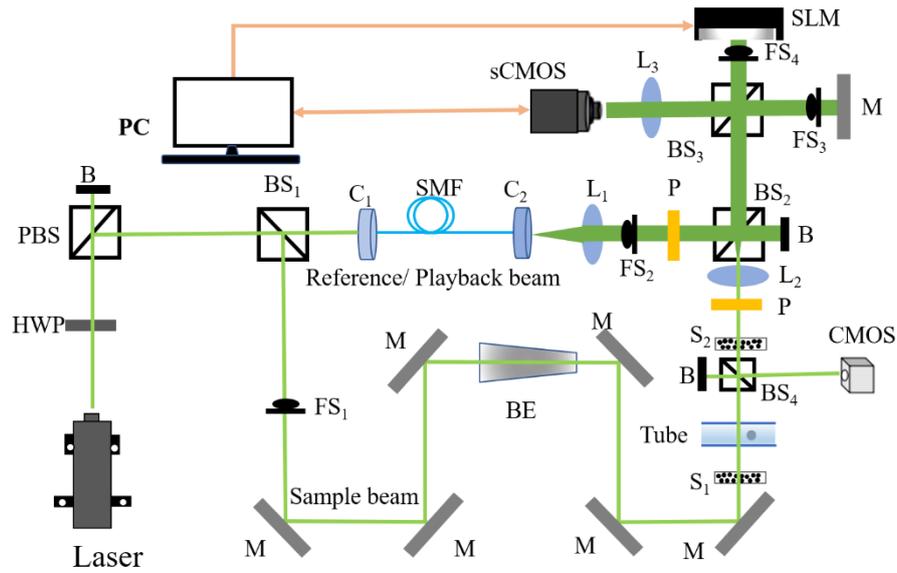

Fig. 1 Schematic of the system. B: beam dump; BE: beam expander; $BS_1$, $BS_2$: cube beam splitter; $BS_3$, $BS_4$: plate beam splitter; $C_1$, $C_2$: fiber port connector; HWP: half-wave plate; $L_1$, $L_2$: Plano-convex lens; $L_3$: camera lens; M: mirror; $FS_1$-$FS_4$: fast shutter; PBS: polarized beam splitter; P: polarizer; $S_1$, $S_2$: scattering layers; sCMOS: scientific CMOS camera; CMOS: CMOS camera; SMF: single mode fiber.

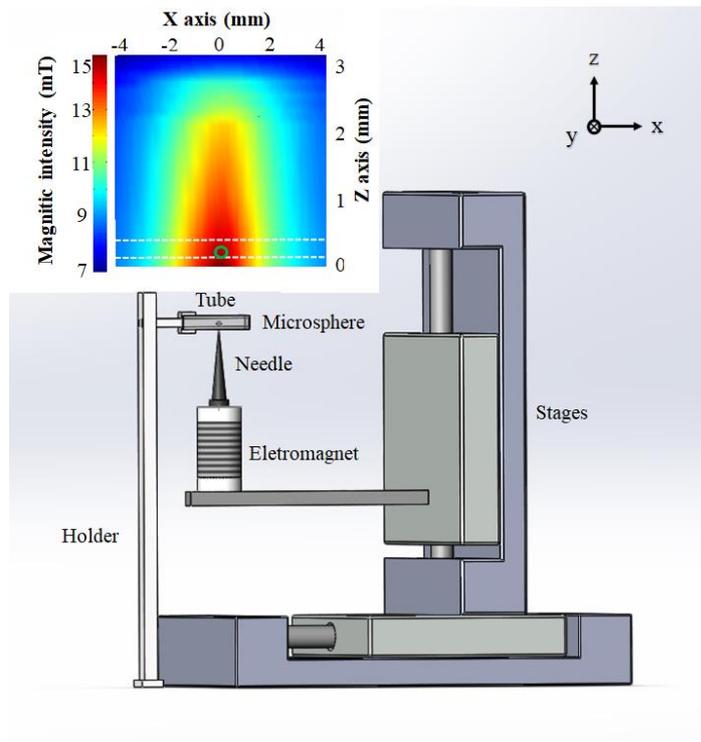

Fig. 2 Illustration of the magnetic control system. Inset: magnetic intensity distribution in the XZ plane above the needle.

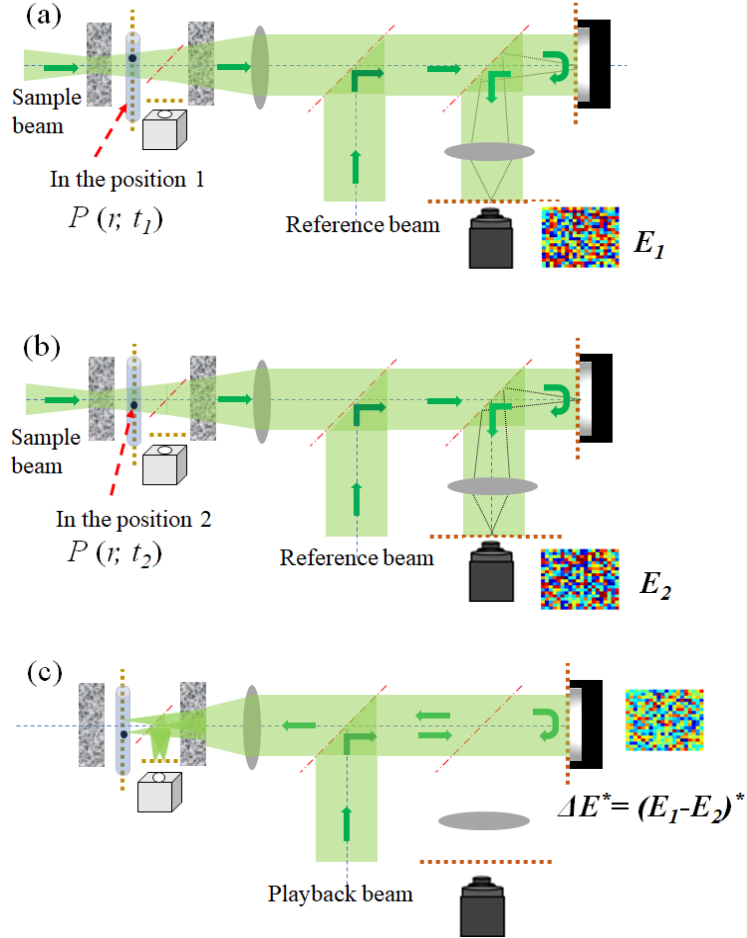

Fig. 3 Illustration of the two-stage TRMCP focusing procedure. (a-b) Hologram recording stage: sample and reference beams interfere, with a camera to record the interference patterns when the magnetically controlled microsphere is at Location 1 and 2, respectively. The Light field difference $ΔE$ is computed and stored. (C) Hologram playback stage: the playback beam (identical to the reference beam) is modulated by the SLM with a phase pattern $ΔE^*$ (the conjugate of $ΔE$), generating a phase conjugation copy of $ΔU$. The new light travels back to the scattering medium, albeit tortuously, and converges to the point of origin—the magnetically controlled moving microsphere.

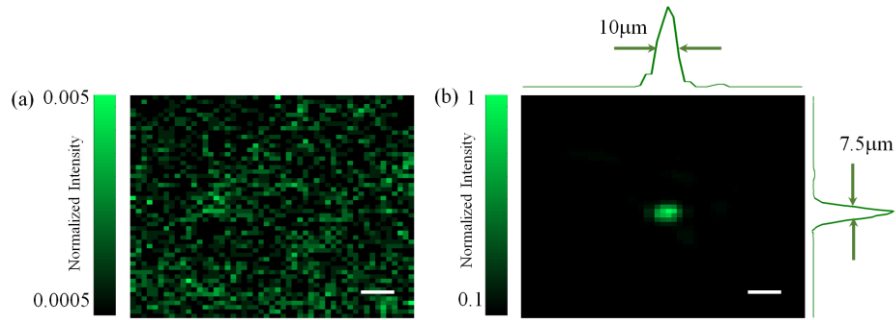

Fig. 4 (a) The optical pattern at the magnetic particle plane between two scattering layers recorded by the CMOS camera when the playback beam illuminates the SLM displayed with a random phase pattern. (b) When the SLM is loaded with the optimized phase pattern, a bright focus is formed with a PBR of 9000. The scalebar represnets 25μm.

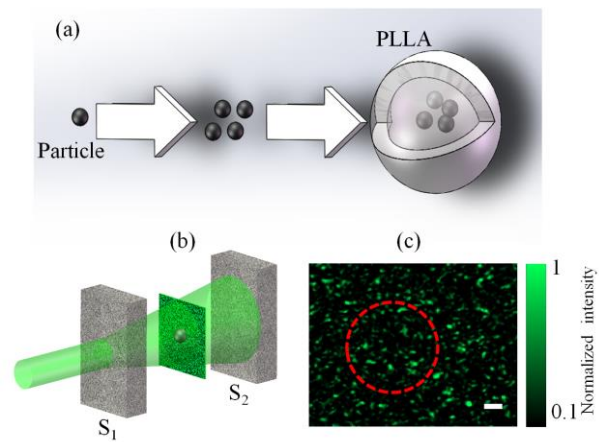

Fig. 5 (a) The enveloping process from magnetic microspheres into a microsphere. (b) The positioning of the magnetic microsphere with respect to the two scattering layers $S_1$ and $S_2$. (c) The resultant random speckle pattern when light illuminates the front surface of $S_1$; the magnetic microsphere is not seen, although a dashed circle is used to indicate its presence. The scalar bar represents 40μm.

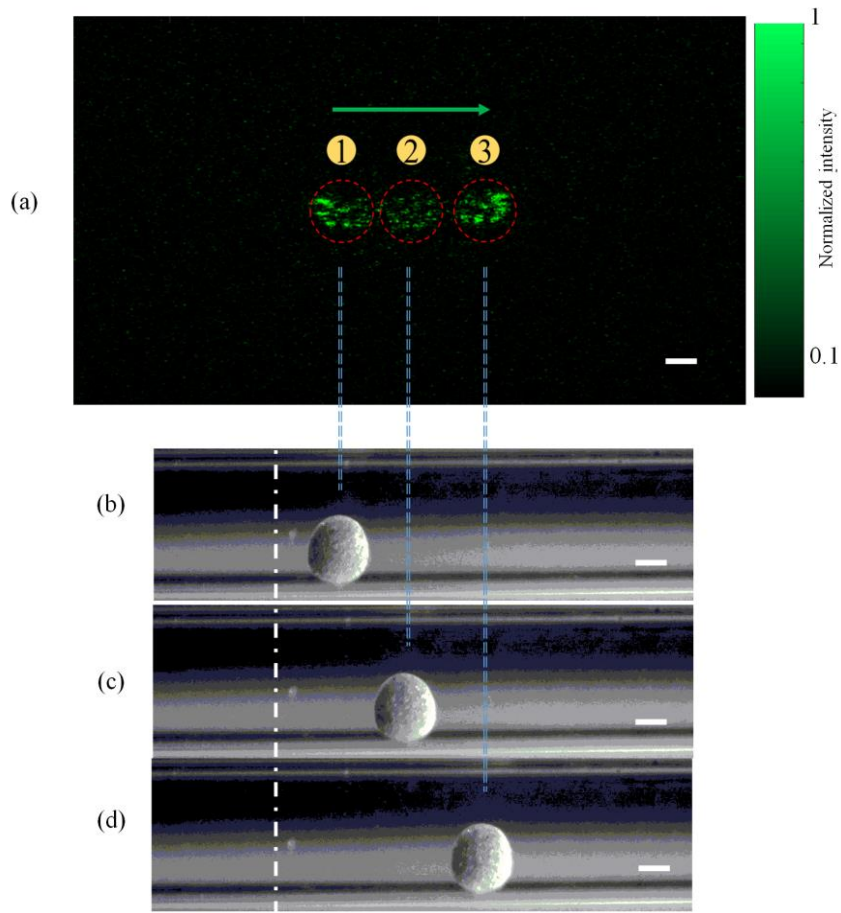

Fig. 6 (a) An illustration of three optical focal spots by using TRMCP, when the magnetic microsphere is externally controlled to move from Location 1 to Location 2, and from Location 2 to Location 3. The dashed circles contour the microsphere region. (b-d) White light images of the microspheres via side detection reveal the microsphere position variation and confirm the performance of TRMCP optical focusing, the scalar bars represent 100μm.